\documentclass[sigconf]{acmart}

\usepackage{enumitem}
\usepackage{color}

\newcommand{\ignore}[1]{}
\newcommand{\an}[1]{$^{_{_{^{^{#1}}}}}$}

\setcopyright{none}
\renewcommand\footnotetextcopyrightpermission[1]{} % removes footnote with conference information in first column

\begin{document}

\title{Significant Improvements over the State of the Art? A Case Study of the MS MARCO Document Ranking Leaderboard}

\author{Jimmy Lin,\an{1,2} Daniel Campos,\an{3} Nick Craswell,\an{2} Bhaskar Mitra,\an{2,4} and Emine Yilmaz\an{4}}

\affiliation{\vspace{0.1cm}
$^1$ University of Waterloo \qquad $^2$ Microsoft AI \& Research \\
$^3$ University of Illinois Urbana-Champaign \qquad  $^4$ University College London 
}

\renewcommand{\shortauthors}{Lin et al.}
\pagestyle{empty}

\begin{abstract}
Leaderboards are a ubiquitous part of modern research in applied machine learning.
By design, they sort entries into some linear order, where the top-scoring entry is recognized as the ``state of the art'' (SOTA).
Due to the rapid progress being made in information retrieval today, particularly with neural models, the top entry in a leaderboard is replaced with some regularity.
These are touted as improvements in the state of the art.
Such pronouncements, however, are almost never qualified with significance testing.
In the context of the MS MARCO document ranking leaderboard, we pose a specific question:\ How do we know if a run is {\it significantly} better than the current SOTA?
We ask this question against the backdrop of recent IR debates on scale types:\ in particular, whether commonly used significance tests are even mathematically permissible.
Recognizing these potential pitfalls in evaluation methodology, our study proposes an evaluation framework that explicitly treats certain outcomes as distinct and avoids aggregating them into a single-point metric.
Empirical analysis of SOTA runs from the MS MARCO document ranking leaderboard reveals insights about how one run can be ``significantly better'' than another that are obscured by the current official evaluation metric (MRR@100).
\end{abstract}

\maketitle

\section{Introduction}

Leaderboard rankings and claims of the ``state of the art'' (SOTA) pervade modern research in applied machine learning, particularly in natural language processing, information retrieval, and computer vision.
There has been much debate in the community on the merits of such activities, compared to alternative uses of the same researcher energy, attention, and resources.
Without participating in this debate, this work attempts to address what we view as a technical shortcoming of many, if not most, leaderboards today:\ the lack of significance testing.

Specifically, we wish to answer the question:\ Does a particular run significantly improve the state of the art?
The nature of leaderboards and rapid progress by researchers mean that the top-scoring run is regularly overtaken and replaced by another run that reports a higher score.
This is communicated (in papers, blog posts, tweets, etc.)\ as beating the existing SOTA and achieving a new SOTA.
Such pronouncements, however, are rarely qualified with significance tests.
We hope to take a small step towards rectifying this.

Our study investigates significance testing among SOTA results in the context of the MS MARCO document ranking leaderboard, but our findings and lessons learned can perhaps be generalized to other leaderboards in IR and beyond.
Against the backdrop of recent debates about evaluation methodology in IR~\cite{ferrante2017ir,sakai2020fuhr,ferrante2021towards}, we discover, unsurprisingly, that there is no simple answer.
Our findings can be summarized as follows:

\begin{enumerate}[leftmargin=*]

\item The existing single-point metric for quantifying the ``goodness'' of a run on the MS MARCO document ranking leaderboard (MRR@100) 
% averages 
%across outcomes that are fundamentally incomparable (with respect to some plausible user model) and this 
conflates important differences in {\it how} one run can be ``better'' than another.
Thus, the naive approach of running standard significance tests on the existing metric may lead to questionable results.

\item To address this issue, we propose an evaluation framework that explicitly tracks outcomes separately, which then permits meaningful aggregation and significance testing.
From a qualitative perspective, this framework reveals many insights about differences that are obscured by the existing official metric.

\item Contributing to recent debates in the IR community on scale types and whether certain statistical operations are mathematically permissible~\cite{ferrante2021towards}, we find that in our framework, analysis in terms of expected search length (ESL), which is a ratio scale, and mean reciprocal rank (MRR), which is an ordinal scale, yield largely the same findings.

\end{enumerate}

\noindent 
The contribution of this work is a novel evaluation framework that compares putative SOTA submissions in a nuanced way that contributes to ongoing debates in the information retrieval community about evaluation methodologies.
We find that runs can be ``better'' in different ways, but these ``different ways'' cannot be reconciled without appealing to a user model of utility (which is presently absent in the task definition).

It is worth emphasizing that in this paper, we are asking a very narrow question about entries on a leaderboard and significance testing with respect to a clearly defined metric (that determines the ranking on the leaderboard).
There are a number of questions that are outside the scope of inquiry, for example:
Is the new SOTA technique practically deployable (e.g., considering inference latencies, model size, etc.)?
Does the model train efficiently? 
Is the improvement in the SOTA meaningful from a user perspective (i.e., does it improve the user experience)?
Might the new SOTA technique encode some bias that would be a cause for concern?
And, no doubt, many more questions.
While these are all important considerations, they raise orthogonal issues that we do not tackle here.
Nevertheless, we show that even for such a narrowly framed question, there is still quite a bit of nuance that is missing in the current discourse. 

\section{Background and Related Work}

The MS MARCO dataset was originally released in 2016 with the aim of helping academic researchers explore information access in the large-data regime, particularly in the context of models based on neural networks that were known to be data hungry~\cite{MS_MARCO_v1}.
Initially, the dataset was designed to study question answering on web passages, but it was later adapted into traditional {\it ad hoc} ranking tasks.
Today, the document ranking and passage ranking tasks host competitive leaderboards that attract much attention from researchers around the world.

This paper focuses on the document ranking task, which is a standard {\it ad hoc} retrieval task over a corpus of 3.2M web pages with URL, title, and body text.
The organizers have made available a training set with 367K queries and a development set with 5193 queries; each query has exactly one relevance judgment.
There are 5793 evaluation (test) queries; relevance judgments for these queries are withheld from the public.
Scores on the evaluation queries can only be obtained by a submission to the leaderboard.
The official metric is mean reciprocal rank at a cutoff of 100 (MRR@100).

Of the myriad metrics that have been proposed to evaluate retrieval systems, there are those that make strong claims as to modeling user utility, such as nDCG~\cite{Jarvelin_Kekalainen_TOIS2002} and RBP~\cite{Moffat_Zobel_2008}, and those that do not, say, precision at a fixed cutoff.
%Note that this is not a binary attribute:\ quite clearly the definition of nDCG, for example, relies heavily on a model of utility.
%However, even metrics we would not usually think of as ``utility-based'' might nevertheless make some plausible claims about modeling utility.
%Take precision at a fixed cutoff $k$:\ this corresponds to a model where the user always examines the top $k$ results with respect to information needs that are ``fairly clear-cut'' (hence, no need for graded relevance judgments); the user gains utility directly proportional to the number of relevant results encountered.
%In the case of factoid question answering on mobile phones, to provide a concrete example, P@3 might not be an unrealistic model of utility.
Specifically in the context of the MS MARCO document ranking task, reciprocal rank (RR) makes at least some plausible claims about utility.
At a high level, the metric says that the user only cares about getting a single relevant document (not unrealistic since MS MARCO models question answering ``in the wild''), and that utility drops off rapidly as a function of increasing ordinal rank.
While the functional form of this dropoff might be a matter of debate (similar disagreements can be had about the functional form of the discount in nDCG), there is strong empirical support for the claim in general, dating back well over a decade.
In the web search context, log analysis (e.g.,~\cite{Agichtein_etal_SIGIR2006a}) as well as eye-tracking experiments (e.g.,~\cite{Joachims05}) have shown that user click probabilities and attention fall rapidly with increasing ordinal rank in the retrieved results.

We believe that at least some of the ongoing controversies about evaluation methodologies in information retrieval stems from confusion on whether a metric is being used simply as a useful proxy for effectiveness (to aid in quantifying model improvements) or is actually making a claim about utility.
Thus, in this paper, we are careful to separate the two, and are explicit when making a claim about utility (and appealing to some user model).

The proximate motivation of this study is the recent work of \citet{ferrante2021towards}, who argued that most IR metrics are not interval-scaled and suggested that decades of IR research may be methodologically flawed and hence the conclusions may not be reliable.
We are not able to provide sufficient justice to their detailed arguments due to space limitations, but the crux of the matter in our context is that for MRR, intervals are not equi-spaced; that is, a difference of $0.1$ (let's say) ``means'' different things at different points on the scale.
As a contrast, the standard example of an interval scale is temperature measured in Celsius, where ``one degree'' means exactly the same thing (i.e., difference in temperature) everywhere.
Recognizing that ongoing debates in IR evaluation methodologies are by no means settled, and without necessarily agreeing with the entirety of their arguments, we take the most conservative position with respect to scale types and permissibility of different operations and statistical analyses. 

\section{Naive First Attempt}
\label{section:first-attempt}

\begin{figure}[t]
\centerline{\includegraphics[width=1.0\columnwidth]{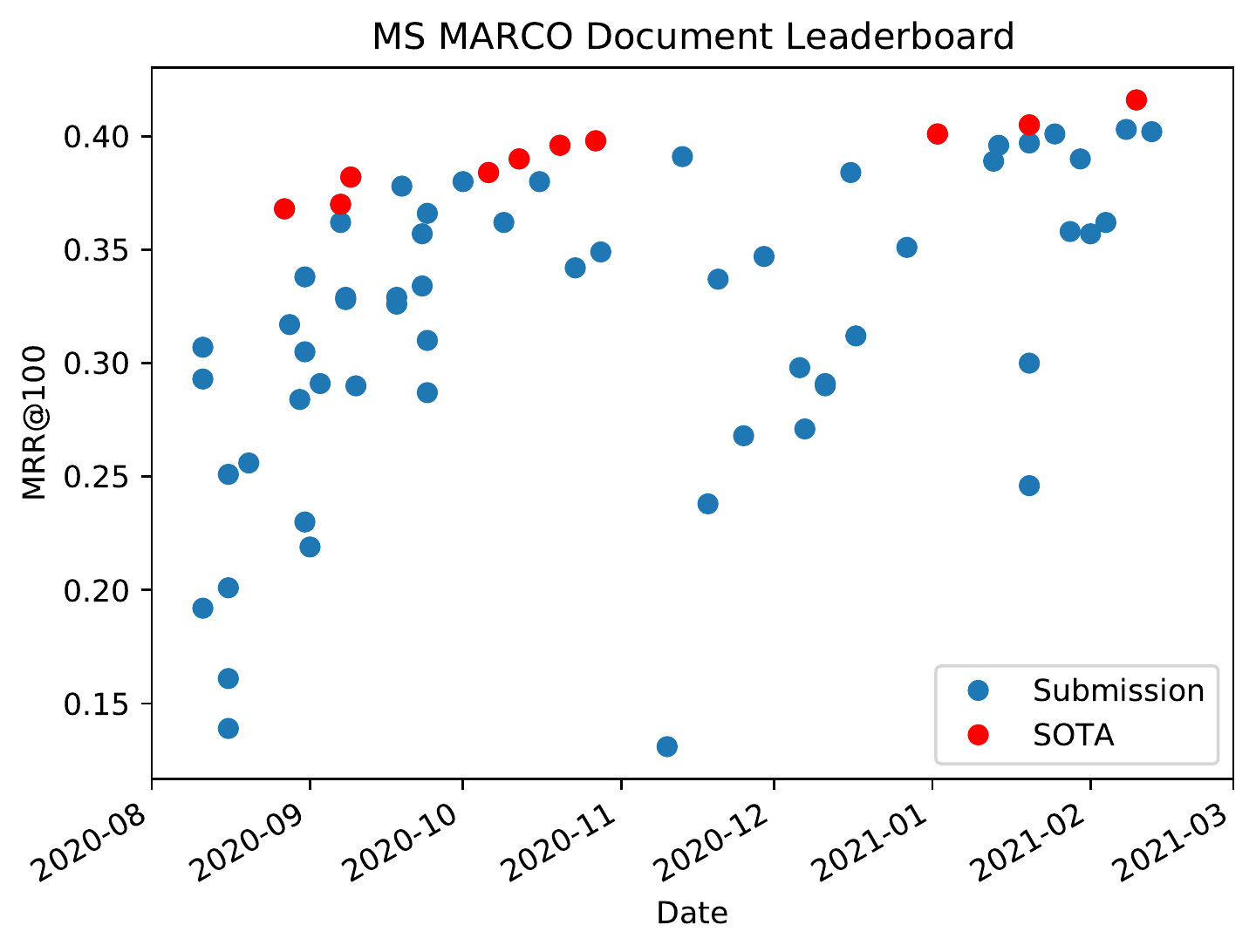}}
\vspace{-0.3cm}
\caption{The leaderboard of the MS MARCO document ranking task, showing the effectiveness of runs (MRR@100) on the held-out evaluation set over time. ``State of the art'' (SOTA) runs are shown in red.} 
\label{fig:overview}
\vspace{-0.3cm}
\end{figure}

A summary of the MS MARCO document ranking leaderboard since its launch in August 2020 (until mid-February, 2021) is shown in Figure~\ref{fig:overview}, where each point represents a run:\ the {\it x}-axis plots the date of submission and the {\it y}-axis plots the official metric (MRR@100) reported on the leaderboard for the held-out evaluation (test) set.
Circles in red represent the (current and former) state-of-the-art (SOTA) runs, i.e., a run that displaced a previous run at the top of the leaderboard, beginning with the first submission that beat the baselines provided by the organizers when the leaderboard first launched.
In our analysis, we specifically focused on these SOTA runs.
Since the identities of the runs (and for that matter, the actual techniques they used) are not germane to our analysis, we simply denote them $R_1$ to $R_{10}$, arranged chronologically.
That is, $R_1$ is the first SOTA run, and $R_{10}$ is the most recent.

If we wish to ask if one SOTA run is significantly better than another, an obvious first attempt would be to run some standard statistical test over per-query scores of the official metric (MRR@100).
Among the myriad tests available, three stand out:

\begin{enumerate}[leftmargin=*]

\item Wilcoxon rank sum test (WRS):\ a non-parametric test that requires samples to be on an ordinal scale.

\item Wilcoxon signed rank test (WSR):\ a non-parametric test that requires samples to be on an interval scale.

\item Student's $t$-test:\ a parametric test that requires samples to be on an interval scale.

\end{enumerate}

\noindent As discussed in \citet{ferrante2021towards}, there has been quite a bit of controversy (in IR and beyond) on what tests are permissible for what scale types.
Even taking the most stringent position, there is no doubt that reciprocal rank is an ordinal scale:\ obtaining a document at rank one is preferable to obtaining a document at rank two, rank three, $\ldots$ is more preferable than not having the document in the ranked list at all.
This holds without making any commitments on the ``distance'' between any of these possible outcomes (i.e., equi-spacing).
Thus, 
%while \citet{ferrante2021towards} would find running WSR and the Student's $t$-test methodologically questionable, 
WRS is unequivocally permissible.
With these caveats rendered explicit, let's just run all the tests anyway.

\begin{table}[t]
\begin{tabular}{llrrrr}
\toprule
\multicolumn{2}{l}{Runs} & &   WRS & WSR & $t$-test \\
$A$ & $B$ & $\Delta$ & $p$-value & $p$-value & $p$-value \\
\cmidrule(lr){1-3} \cmidrule(lr){4-6} 
$R_1$ & $R_2$ & 0.0027 & 0.029393 &	0.519758	& 0.722586 \\
$R_1$ & $R_3$ & 0.0152 & 0.109672 & 0.053127	& 0.038240 \\
$R_1$ & $R_4$ & 0.0179 & 0.000494 &	0.008780	& 0.013607 \\
$R_1$ & $R_5$ & 0.0244 & 2.07E-05 &	0.000461	& 0.000785 \\
$R_1$ & $R_6$ & 0.0301 & 1.16E-06 &	3.21E-05	& 3.77E-05 \\
$R_1$ & $R_7$ & 0.0328 & 1.16E-07 &	1.00E-06	& 4.18E-06 \\
$R_1$ & $R_8$ & 0.0360 & 2.81E-05 &	3.24E-06	& 2.50E-06 \\
$R_1$ & $R_9$ & 0.0406 & 4.15E-06 &	9.06E-08	& 3.61E-08 \\
$R_1$ & $R_{10}$ & 0.0516 & 7.85E-14 & 1.92E-12	& 7.57E-12 \\
\midrule
$R_9$ & $R_{10}$ &0.0110	&	0.010663 &	0.109114 & 0.153225 \\
\bottomrule
\end{tabular}
\vspace{0.25cm}
\caption{Results of running significance tests on SOTA runs:\ Wilcoxon rank sum test (WRS), Wilcoxon signed rank test (WSR), and the Student's $t$-test. $R_1 \ldots R_{10}$ are the SOTA runs, arranged chronologically.}
\label{result:naive}
\vspace{-0.4cm}
\end{table}

The results of running all three significance tests on pairwise comparisons between $R_1$ (the earliest SOTA run) and every other subsequent SOTA run $\{ R_2 \ldots R_{10} \}$ on the evaluation set are shown in Table~\ref{result:naive}.
Additionally, we compare the two current top runs on the leaderboard, $R_9$ and $R_{10}$.
The table reports the absolute differences in effectiveness,\footnote{There is an important detail here worth mentioning:\ the official evaluation script deliberately introduces a metric artifact designed to thwart (simple) attempts at ``reverse-engineering'' the evaluation set. Thus, the scores reported on the official leaderboard (and plotted in Figure~\ref{fig:overview}) are {\it not} accurate. This artifact has no impact on the leaderboard rankings, but {\it does} impact significance testing. All our analyses are based the true MRR@100 scores, after the removal of this artifact. Thus, the absolute score differences may not line up with public leaderboard results.} along with the raw $p$-values of the different tests, prior to the application of the Bonferroni correction for multiple hypothesis testing.

We see that based on all three tests, the improvements from successive SOTA runs $R_2, R_3, \ldots$ are not statistically significant until we get to $R_5$; all subsequent runs thereafter appear to be significantly better (even just focusing on WRS).
The absolute difference in MRR@100 between $R_1$ and $R_5$ is 2.4 points, which is surprisingly large.
Independent of the particulars of any evaluation, the general expectation would be that with a large number of queries (over 5K in our case), small significant differences (i.e., small effect sizes) should be detectable.
That doesn't appear to be the case here.
While this unexpected finding does not in and of itself demonstrate that something is amiss, it would be natural to suspect Type II errors---that is, our tests are not powerful enough to confidently reject the null hypothesis.
As it turns out, this is not the issue.
We will demonstrate in the context of our proposed evaluation framework:\ the fundamental issue is that MRR@100 forces aggregation across outcomes that are not directly comparable (with respect to some plausible user model).

\section{Outcomes Breakdown}

This section presents our evaluation framework specifically tailored to the MS MARCO document ranking task.
We begin with a few (hopefully) uncontroversial claims and from there build an approach to evaluation that explicitly avoids conflating distinct outcomes in a single-point metric.
Note that since the official relevance judgments contain only one relevant document per query, the position of that relevant document on the ranked list (or its absence) alone determines the score (the metric, to be defined below) for that query; this nicely sidesteps the challenges with different ``recall bases''~\cite{ferrante2021towards}, i.e., queries that have different numbers of relevant documents.

Consider two hypothetical submissions to the MS MARCO document ranking leaderboard, runs $A$ and $B$, comprising ranked lists over a set of queries $Q$.
For each query $q \in Q$, there are logically the following distinct outcomes that cover all possibilities:

\begin{enumerate}[leftmargin=*]

\item Neither run $A$ nor run $B$ returns the relevant document in the top $k$ ranking.
In this case, both runs are equally ``bad''.

\item Run $A$ returns the relevant document in the top $k$ ranking, while run $B$ does not. Run $A$ is thus ``better''. Vice versa with $B$ and $A$ swapped.

\item Both run $A$ and $B$ return the relevant document in the top $k$ ranking, but the document has a ranking closer to the front of the ranked list in run $B$ (i.e., lower ordinal rank). Run $B$ is thus ``better''. Vice versa with $B$ and $A$ swapped.

\end{enumerate}

\noindent We believe that the above assertions hold regardless of how one might choose to operationalize ``bad'' and ``better''.
However, to be more precise, let us define a metric in terms of expected search length (ESL), which has a long history in IR research dating back to the 1960s~\cite{ESL}.
ESL quantifies how long a user needs to search (more specifically, read the ranked list) before obtaining a relevant document:
A relevant document appearing at rank 1 gets a score of 1, rank 2 gets a score of 2, etc.\ all the way up to rank 100 (in our case).
Thus, the lower the score, the better.

Consider a straightforward user model, that of a patient user who issues a query and is willing to read 100 documents per query (at a constant pace) to find the relevant document, and then gives up (if no relevant document is found).
It would be plausible to make the claim that ESL, with respect to this user model, captures utility measured in user time.\footnote{Recognizing that we are making a few simplifying assumptions such as constant document length and fixed reading speed. More realism could be added by, for example, taking into account a more accurate model of reading speed~\cite{Smucker_Clarke_SIGIR2012}, but these refinements are unlikely to change our overall analysis.}

\begin{table*}[t]
\begin{tabular}{llrrrrrrrrrrrrr}
\toprule
\multicolumn{2}{l}{Runs} & & \multicolumn{1}{l}{\bf (1)} & \multicolumn{1}{l}{\bf (2)} &  & \multicolumn{1}{l}{\bf (3)} & & & WSR & $t$-test & & & WSR & $t$-test \\
$A$ & $B$ & $\Delta$ & All & $A$ wins & $B$ wins & All & $A$ ESL & $B$ ESL & $p$-value & $p$-value & $A$ RR & $B$ RR & $p$-value & $p$-value \\
\cmidrule(lr){1-3} \cmidrule(lr){4-4} \cmidrule(lr){5-6} \cmidrule(lr){7-7} \cmidrule(lr){8-11} \cmidrule(lr){12-15}
$R_1$ & $R_2$ & 0.0027 & 2\% & 9\%  & 16\% & 73\% & 6.99 & 9.14 & 3.22E-12	& 1.60E-07 &0.4812	&0.4468	&6.75E-05	&2.66E-05 \\
$R_1$ & $R_3$ & 0.0152 & 4\% & 15\% & 14\% & 67\% & 7.15 & 5.99 & 4.62E-06	& 1.34E-05 &0.4857	&0.5107	&0.004032	&0.003871\\
$R_1$ & $R_4$ & 0.0179 & 3\% & 10\% & 15\% & 72\% & 7.06 & 8.13 & 0.000249 & 0.004918 &0.4838	&0.4775	&0.448624	&0.368583\\
$R_1$ & $R_5$ & 0.0244 & 3\% & 11\% & 15\% & 71\% & 7.13 & 7.27 & 0.609428 & 0.813828 &0.4833	&0.4861	&0.730242	&0.765044\\
$R_1$ & $R_6$ & 0.0301 & 3\% & 10\% & 15\% & 71\% & 7.15 & 7.19 & 0.886547 & 0.381946 &0.4829	&0.4910	&0.331145	&0.364266\\
$R_1$ & $R_7$ & 0.0328 & 4\% & 10\% & 15\% & 72\% & 7.27 & 6.98 & 0.279188 & 0.022938 &0.4799	&0.4979	&0.024974	&0.020952\\
$R_1$ & $R_8$ & 0.0360 & 3\% & 15\% & 15\% & 67\% & 7.06 & 5.71 & 7.94E-08 & 8.68E-08 &0.4829	&0.5273	&8.97E-07	&7.59E-07\\
$R_1$ & $R_9$ & 0.0406 & 4\% & 15\% & 14\% & 67\% & 7.15 & 5.15 & 2.03E-16 & 4.33E-17 &0.4857	&0.5404	&2.49E-10	&1.85E-10\\
$R_1$ & $R_{10}$ & 0.0516 & 3\% & 12\% & 15\% & 70\% & 7.08 & 5.88 & 4.79E-06 &	1.46E-06 &0.4829	&0.5204	&1.60E-05	&1.10E-05\\
\midrule
$R_9$ & $R_{10}$ & 0.0110 & 3\% & 12\% & 17\% & 69\% & 5.22	&5.79	& 0.026583 &	0.006238 &	0.5376	& 0.5222	& 0.082654 &	0.077858 \\
\bottomrule
\end{tabular}
\vspace{0.25cm}
\caption{Analysis of SOTA runs from the MS MARCO document ranking leaderboard, broken into distinct outcomes:\ (1) neither run retrieves the relevant document, (2) one run retrieves the relevant document but not the other, and (3) both runs retrieve the relevant document.} 
\label{result:msmarco-doc}
\vspace{-0.4cm}
\end{table*}

It is clear that ESL as articulated here is on a ratio scale (which is by definition also an interval scale).
A difference of 1 ``means the same thing'' at 2 ESL as it does at 99 ESL:\ in both cases, the relevant document appears one rank closer to the top of the ranked list; from the utility perspective, in both cases the user has saved the same amount of time.\footnote{Once again, setting aside refinements that better model document length, reading speed, etc.}
Furthermore, being a ratio scale, 3 ESL is 3$\times$ worse than 1 ESL, and 100 ESL is 5$\times$ worse than 20 ESL (the analogy here is that ESL is comparable measuring temperature in Kelvins).
Against our user model, these are also plausible statements to make with respect to utility (time):\ a relevant document at rank 4 (4 ESL) costs the user twice as much utility (time) as a relevant document at rank 2 (2 ESL).

So, for case (3) in the list of outcomes above when comparing run $A$ and run $B$ for a specific query $q$, we have a good alignment between ESL and utility, and we can make a number of meaningful claims.
For example, the following statements are permissible and indeed meaningful:

\begin{itemize}[leftmargin=*]

\item For $q_1$, run $A$ returns the relevant document at rank 2 and run $B$ returns the relevant document at rank 4.
For $q_2$, run $A$ returns the relevant document at rank 50 and run $B$ returns the relevant document at rank 52.
In both cases, $A$ is better than $B$ by two ``ESL units'', and even more strongly, two units of utility (time).

\item  For $q_3$, run $A$ returns the relevant document at rank 1 and run $B$ returns the relevant document at rank 4.
For $q_4$, run $A$ returns the relevant document at rank 10 and run $B$ returns the relevant document at rank 40.
In both cases, $A$ is better than $B$ by 4$\times$, in terms of both ``ESL units'' and units of utility.

\end{itemize}

\noindent The upshot here is that we can meaningfully average ESL values across a large set of queries $Q$.
Consider:\ for $q_5$, run $A$ returns the relevant document at rank 1 and run $B$ returns the relevant document at rank 4, and for $q_6$, run $A$ returns the relevant document at rank 5 and run $B$ returns the relevant document at rank 2.
We can say that on the set of queries comprised of $\{ q_5, q_6\}$ both runs are equally effective, with an (arithmetic) mean ESL of 3.
The arithmetic mean here encodes the assumption that each query is equally important, which seems reasonable.
Furthermore, it is plausible to say that on these two queries, a user derives equal utility.

Note, critically, however, that this only applies to case (3) above, when both runs contain the relevant document in their top $k$ lists.
For case (2), however, it is unclear how similar statements can be made.
Consider:\ for $q_7$, run $A$ returns the relevant document at rank 1 and run $B$ returns the relevant document at rank 100, and for $q_8$, run $A$ doesn't return the relevant document at all and $B$ returns the relevant document at rank 99.
There is little that we can meaningfully say about the effectiveness of the runs on a set of queries comprised of $\{ q_7, q_8\}$, both from the perspective of ESL (what ESL would we assign to run $A$ for $q_8$?)\ and utility.
For the latter, we would need to quantify the cost of not finding a relevant document relative to an ESL or time unit.
There's nothing in the framework we've presented thus far that would shed light on this without a more articulated user model (absent in the current task definition).
Note that the official metric MRR@100 {\it does} encode a specific utility difference between a retrieved document at rank 100 and not retrieving the relevant document ($0.01$ and $0$, respectively), but justifying these values would require appeal to user models and data (e.g., query logs) that are beyond the scope of the leaderboard.
We argue, instead, that the best way forward is to maintain an explicit separation and breakdown of the different outcomes.
In other words, cases (2) and (3) represent apples and oranges, and it would be suspect to average across them without first establishing some way to compare different fruits.

\section{Application to MS MARCO}

Let us now apply the evaluation framework proposed above to analyze the SOTA runs on the MS MARCO document ranking leaderboard.
In our analysis, we compared each of $R_2 \ldots R_{10}$ against $R_1$, the results of which are shown in Table~\ref{result:msmarco-doc}; we additionally compared the top two runs on the leaderboard, $R_{10}$ against $R_9$.
We show the percentage of queries that fall under a particular outcome---case (1), (2), or (3)---as described in the previous section.
Case (2) is broken down into ``$A$ wins'', where $A$ returns the relevant document in its ranked list and $B$ doesn't, and ``$B$ wins'', the opposite case.
For rhetorical convenience, we will use ``answered'' and ``unanswered'' for these cases.
For case (3), we show the overall percentage, as well as the mean ESL and reciprocal rank (RR) for all queries that fall under that outcome.
For ESL and RR, we show $p$-values from the application of the Wilcoxon signed-rank test and the paired $t$-test.
Note that since ESL is on an interval (ratio) scale, these two tests are unequivocally permissible.
For RR, the application of the Wilcoxon signed-rank test and the paired $t$-test is subjected to the potential objections raised by \citet{ferrante2021towards} regarding interval scales.
In all cases, we report raw $p$-values, prior to the application of the Bonferroni correction for multiple hypothesis testing.

This specific case study, and more generally our proposed evaluation framework, reveals many interesting insights that are completely hidden if we simply reported the means of per-query reciprocal ranks and ran significance tests on them, as in Section~\ref{section:first-attempt}.
Focusing specifically on ESL, we highlight a number of interesting observations below:

\begin{itemize}[leftmargin=*]

\item In the first row, comparing $R_1$ vs.\ $R_2$, we see the overall MRR@100 scores are quite close, but both runs obtain the scores in very different ways.
Looking at the case (3) breakdowns, we see that $R_2$ has a higher ESL than $R_1$, and this difference is (highly) statistically significant.
From this perspective, $R_2$ is worse than $R_1$ (relevant results appearing later in the ranked lists).
However, we see that $R_2$ answered far more queries that went unanswered in $R_1$ than the other way around.

\vspace{0.05cm}
A similar observation can be made in the comparison between $R_1$ and $R_4$.
When focusing only on ranking, case (3), $R_4$ is significantly worse than $R_1$, but $R_4$ compensates by answering more queries that went unanswered in $R_1$, leading to a higher score in terms of MRR@100.

\vspace{0.05cm}

\item Consider the comparison between $R_1$ and $R_3$ (the second row):\ contrary to the examples discussed above, we see that $R_3$ significantly improves ranking, case (3), but has slightly more unanswered queries compared to the baseline.
This also leads to an overall improvement in terms of MRR@100.

\vspace{0.05cm}

\item Consider the comparison between $R_1$ and\ $R_8$.
Here, looking at the prevalence of case (2):\ there are equal percentages of cases where the query was answered by one run but not the other.
However, looking at case (3), we see that $R_8$ obtains a statistically significant reduction in ESL.
That is, $R_8$ is better than $R_1$ not because it obtained more relevant documents, but rather that it ranked the relevant documents more highly.

\vspace{0.05cm}

\item Another interesting observation relates to $R_1, R_5, R_6, R_7$, which are runs from the same team.
From the rows comparing $R_1$ to $\{ R_5, R_6, R_7 \}$, we see that in terms of case (3), differences in ESL are not statistically significant.
That is, the runs are comparable when it comes to ranking documents that appear in the top 100.
The differences in MRR@100 come primarily from case (2), where $\{ R_5, R_6, R_7 \}$ have fewer unanswered queries on balance.

\vspace{0.05cm}

\item Looking at the current top two runs on the leaderboard, $R_{10}$ and $R_9$, we see that the second-best run actually does a better job ranking, i.e., case (3), than the top run.
The latter wins by virtue of it answering more queries.

\end{itemize}

\smallskip
\noindent
Examining the differences between ESL and RR, our analysis shows that they largely agree, with the exception of $R_1$ vs.\ $R_3$ and $R_1$ vs.\ $R_4$.
It seems that differences in ESL are relatively larger than differences in RR, which could explain these cases.
Note that two runs with the same ESL can have very different MRRs.
For example, consider the case of a run retrieving relevant documents at ranks 1 and 9, and a second run retrieving relevant documents at ranks 4 and 6.
Both runs would have the same ESL, but very different MRRs, 0.556 and 0.208, respectively.
There may be a tempting intuitive interpretation of MRR as the reciprocal of the mean position at which a relevant document appears, but this interpretation is not accurate.

With respect to the ongoing IR debate on scale types, the conclusion seems to be that, at least in our proposed framework, {\it it doesn't matter much}.
That is, even adopting the most cautious position advocated by~\citet{ferrante2021towards}, analysis in terms of ESL (a ratio scale) and MRR ({\it not} an interval scale) does not lead to markedly different conclusions.
Given this, there may be reasons to prefer RR over ESL, since the former captures a more realistic user model; it is highly unlikely that user attention can be sustained across the examination of 100 results, as suggested by ESL.
Here, the theory of IR evaluation appears to diverge from its actual practice on large-scale, real-world datasets.

Back to the original question that we set out to answer:\ Is ``this'' SOTA run better than ``that'' SOTA run?
We might say that run $B$ is better than run $A$ if run $B$ wins in terms of case (2) {\it and} has a smaller ESL for case (3), or alternatively, greater MRR.
We might further claim that run $B$ is significantly better than run $A$ if the improvements in both outcomes are statistically significant:\ for case (3), significant testing as we have performed above, and for case (2), perhaps the binomial test.

Alternatively, we might adopt a less stringent definition, somewhat akin to the Hippocratic Oath (i.e., ``do no harm''):\ a run can be considered significantly better if it significantly increases the fraction of answered questions without significantly increasing the ESL {\it or} that it significantly decreases ESL without significantly increasing the fraction of unanswered questions.
The advantage of this approach is that it provides two concrete facets of ``goodness'' that researchers can independently tackle while still being amenable to a linear sort order for populating a leaderboard.

\section{Concluding Thoughts}

One attraction of leaderboards is that they define a concrete metric to optimize, and if that metric is well-defined and meaningful, it enables the community to make rapid progress on a problem.
This work focuses on one important yet under-explored aspect of intrinsic validity:\
We show that naive significance testing applied to MRR@100 obscures many potential insights, and that our proposed framework provides a more nuanced analysis.

How might we generalize our framework to encompass other retrieval tasks and  possibly beyond?
We see at least one challenge that limits the broader applicability of our approach:\ it critically depends on having only one relevant document per query, since this property is necessary to separate the outcomes.
Otherwise, it is unclear how we would compare ranked lists that retrieve different numbers of relevant documents.
While this restriction would not be unrealistic for some tasks, such as question answering, there clearly needs to be more work before our approach can be generalized.

\bibliographystyle{ACM-Reference-Format}
\bibliography{sota}

\end{document}